\documentstyle[aps]{revtex}

\begin{document}
\draft
\author{Alain F. Veilleux, Anne-Marie Dar\'{e}, Liang Chen, Y.M.Vilk,
and A.-M.S.
Tremblay}
\title{Magnetic and pair correlations of the Hubbard model with
next-nearest-neighbor hopping. }
\address{D\'{e}partement de physique and Centre de recherche
en physique du solide.\\
Universit\'{e} de Sherbrooke, Sherbrooke, Qu\'{e}bec, Canada J1K 2R1}
\date{13 July 1995, cond-mat/9507045 }
\maketitle

\begin{abstract}
A combination of analytical approaches and quantum Monte Carlo simulations
is used to study both magnetic and pairing correlations for a version of the
Hubbard model that includes second-neighbor hopping $t^{\prime }=-0.35t$ as
a model for high-temperature superconductors. Magnetic properties are
analyzed using the Two-Particle Self-Consistent approach. The maximum in
magnetic susceptibility as a function of doping appears both at finite $%
t^{\prime }$ and at $t^{\prime }=0$ but for two totally different physical
reasons. When $t^{\prime }=0$, it is induced by antiferromagnetic
correlations while at $t^{\prime }=-0.35t$ it is a band structure effect
amplified by interactions. Finally, pairing fluctuations are compared with $%
T $-matrix results to disentangle the effects of van Hove singularity and of
nesting on superconducting correlations. The addition of antiferromagnetic
fluctuations increases slightly the $d$-wave superconducting correlations
despite the presence of a van Hove singularity which tends to decrease them
in the repulsive model. Some aspects of the phase diagram and some
subtleties of finite-size scaling in Monte Carlo simulations, such as
inverted finite-size dependence, are also discussed.
\end{abstract}


\section{Introduction}

High-temperature superconductors pose a theoretical challenge\cite{Revue}
for many reasons. One of them is that their band structure does not suffice
by itself to explain most experiments, even in the normal state. This is one
of the reasons that much work has been devoted to the simplest model which
incorporates the effects of short-range interactions, the Hubbard model,

\begin{equation}
H=-\sum_{<ij>\sigma }t_{i,j}\left( c_{i\sigma }^{\dagger }c_{j\sigma
}+c_{j\sigma }^{\dagger }c_{i\sigma }\right) +U\sum_in_{i\uparrow
}n_{i\downarrow \,\,\,\,}.  \label{Hubbard}
\end{equation}
In this expression, the operator $c_{i\sigma }$ destroys an electron of spin
$\sigma $ at site $i$. Its adjoint $c_{i\sigma }^{\dagger }$ creates an
electron. The symmetric hopping matrix $t_{i,j}$ determines the band
structure. We consider the case where $t_{i,j}=t$ for nearest-neighbors, $%
t_{i,j}=t^{\prime }$ for next-nearest neighbors and $t_{i,j}=0$ otherwise.
Double occupation of a site costs an energy $U$ due to the screened Coulomb
interaction. We work in units where the lattice spacing is unity, $k_B=1$, $%
\hbar =1$ and $t=1$. Since high-temperature superconductivity generally
occurs in systems with a pronounced planar structure, the two-dimensional
square lattice version of the model is considered, as usual.

The nearest-neighbor $\left( t^{\prime }=0\right) $ version of the above
Hubbard model has been extensively studied. Within this model, the only hope
to get even qualitative agreement with experiment is if the interactions or
the antiferromagnetic fluctuations are very strong. Indeed, with the
nearest-neighbor model the Fermi surface topology and the filling dependence
of both the Hall coefficient and the uniform magnetic susceptibility are all
qualitatively wrong, a disagreement that cannot be removed perturbatively.
On the other hand with second nearest-neighbor hopping, the band structure
becomes more realistic and all the above physical quantities as well as the
position of neutron scattering intensity maxima\cite{Levin}\cite{Littlewood}%
\cite{Lavagna}\cite{Benard} have at least the correct qualitative behavior.

It is thus important to understand the Hubbard model Eq.(\ref{Hubbard})
including both nearest-neighbor hopping $t$ and next-nearest-neighbor
hopping $t^{\prime }$ since weaker interactions in this more realistic model
might suffice to explain the experimental data. Two main questions are
addressed here by a combination of numerical\cite{Moreo}\cite{LinHirsch}\cite
{Lin}\cite{DosSantos}\cite{Husslein}\cite{Imada}\cite{Duffy} and analytical
approaches:\cite{Veilleux} First, the origin of the maximum in the uniform
magnetic susceptibility as a function of band filling at low temperature,
and second the relative influence of density of state effects (van Hove
singularity) and Fermi surface-topology effects (nesting) on pairing
correlations. While both van Hove singularity and nesting occur
simultaneously at half-filling in the usual $t^{\prime }=0$ model, there is
no nesting when $t^{\prime }\neq 0$.\cite{TheseBenard}

In the rest of this paper we first discuss the Monte Carlo approach and the
range of parameters studied. In this same section, we discuss subtleties of
finite-size scaling including a new effect, inverted finite-size scaling. In
the following two sections, we present results in turn for the magnetic
susceptibility and for antiferromagnetic as well as pairing correlations
along with analytical\cite{Note} approaches to understand the results. It
will be shown that for magnetic properties, the recently proposed
Two-Particle Self-Consistent (TPSC) approach\cite{Vilk} compares very well
with Monte Carlo results, allowing us to understand the origin of the
maximum in magnetic uniform magnetic susceptibility as well as some features
of the phase diagram. The same kind of agreement was found before in the
case of the nearest-neighbor model.\cite{Vilk} Superconducting fluctuations
on the other hand are accounted for by $T$-matrix effects far from the van
Hove singularity \cite{Dare}\cite{DareThese}. The physical interpretation of
the results is summarized in the conclusion.

\section{Quantum Monte Carlo approach, simulation parameters and finite-size
effects}

After a brief discussion of technical details of the simulations, we make
several general points about finite-size effects. These effects can be
especially unusual in the presence of $t^{\prime }$. We point out in
particular how false signals of antiferromagnetism can be detected as an
inverse finite-size dependence. Finite-size effects will be discussed
further in the other sections for each specific physical situation
encountered.

\subsection{Simulations}

We use the so-called determinantal quantum Monte Carlo approach\cite{Hirsch}%
\cite{BSS}\cite{BSSRevue}\cite{White}. As usual the sign of the fermion
determinant renders the statistical accuracy very poor in certain regions of
parameter space so that long computation times are required. Figure\ 1,
computed using the standard gauge\cite{ChenJauge} for the
Hirsch-Hubbard-Stratonovich transformation, gives the average sign of the
fermion determinant for typical parameters considered in the present paper.
For $t^{\prime }=0$, the sign was always positive at half-filling because of
particle-hole symmetry\cite{HirschVieux} but this is no longer the case for $%
t^{\prime }\neq 0$.

All the simulations presented below are for either $t^{\prime }=0$, or $%
t^{\prime }=-0.35$ which, in addition to minimizing finite-size effects, as
described below, is typical for high-temperature superconductors. Indeed,
band-structure calculations\cite{Hybertsen}\cite{Pickett} suggest $t^{\prime
}=-0.16$ for $La_2CuO_4$ and $t^{\prime }=-0.45$ for $YBa_2Cu_3O_7.$

Two to five different runs were done for each point with $6\times 10^5$ to $%
2\times 10^6$ measurements per run. Measurements were grouped into $400$
equal-size blocks to estimate the statistical error. Between $10^{3\text{ }}$%
and $10^4$ flips per spin in space-time were performed for the warm-up
period. Sparse matrix techniques were used for the exponential of the
kinetic energy operator.\cite{White} The most difficult points took 300
hours each to be obtained at a computing speed of 26 million floating point
operations per second. The imaginary-time discretization step was between $%
1/8$ and $1/16$. Some results were also obtained on a Fujitsu supercomputer.

\subsection{Finite-size effects}

One should recall that finite-size corrections arise at low temperature
because of two different physical effects: a) The correlation length arising
from interaction-induced collective phenomena (antiferromagnetism) can
become larger than the system size. b) The temperature can become smaller
than the energy-level separation or, equivalently, the thermal de Broglie
wavelength $\xi _{th}=v_F/\left( \pi T\right) $ can become larger than the
system size. In the latter case, even the results for $U=0$ contain
finite-size effects.

Since finite-size effects limit the lowest temperature that we can consider,
they also determine the smallest accessible value of the next-nearest
neighbor hopping $t^{\prime }$. Indeed, non-trivial effects caused by $%
t^{\prime }$ appear only when $t^{\prime }>T$. Furthermore, it is important
to realize that finite-size corrections may depend on filling at fixed
temperature. This is particularly important in the context of one of our
problems, namely the maximum in the uniform magnetic susceptibility as a
function of doping at low temperature. Our results, in Figs.2 and 3, will be
discussed in more detail in the next section. To understand finite-size
effects here, we first recall that the uniform magnetic susceptibility of
the non-interacting infinite system has a maximum away from half-filling as
long as $t^{\prime }\neq 0$. In a $10\times 10$ system at $U=0$, $\beta
=1/T=5$, one can check that this maximum first appears away from
half-filling only for $t^{\prime }$ in the vicinity of $-0.35.$ Smaller
values of $t^{\prime }$ do not even lead to a filling dependence
qualitatively similar to that of the infinite system at zero-temperature.
For $t^{\prime }=-0.35$, several of the allowed wave vectors of the finite
system are close to the Fermi surface at the filling $n$ about $0.6$ where
is located the van Hove singularity in the infinite-size system. This leads
to a better finite-size representation of this singularity. The finite-size
effects in the non-interacting $8\times 8$ system illustrated by the dashed
line in Fig.3 lead to the small oscillations around the infinite-size
result. These oscillations would become much larger at lower temperature
where they could lead to erroneous conclusions.

The interplay between finite-size effects and antiferromagnetic fluctuations
in the finite $t^{\prime }$ model is also very peculiar. Antiferromagnetic
fluctuations may be observed in the $\left( \pi ,\pi \right) $ component of
the magnetic structure factor defined by,
\[
S\left( {\bf q}\right) =\frac 1N\sum_{{\bf r}_i,{\bf r}_j}\exp \left[ i{\bf %
q\cdot }\left( {\bf r}_i-{\bf r}_j\right) \right] \left\langle \left(
n_{i,\uparrow }-n_{i,\downarrow }\right) \left( n_{j,\uparrow
}-n_{j,\downarrow }\right) \right\rangle .
\]
Fig. 4 allows one to suggest that in the $t^{\prime }=-0.35$ model the
antiferromagnetic fluctuations are largest around $n=1$ since this is where
finite-size effects are largest. By contrast with the $t^{\prime }=0$ model
however, we observe here an {\it inverted finite-size effect}. What we mean
is that the magnetic structure factor is converging {\it down} to the
infinite volume result. In the $t^{\prime }=0$ model, this is never
observed. When the correlation length is finite but larger than the system
size at this temperature, $S\left( {\bf q}\right) $ normally {\it increases}
with system size until it saturates. The TPSC described below\cite{Vilk}
reproduces the inverted finite-size effect of the $t^{\prime }=-0.35$ model
and suggests that Monte Carlo calculations on $10\times 10$ would be
necessary to converge to the infinite-size result at this temperature. The
source of the inverted finite-size effect can be found already from the $U=0$
calculation which shows that for $t^{\prime }=-0.35$ the magnitude of the $%
{\bf q=}\left( \pi ,\pi \right) $ susceptibility on a $6\times 6$ or a $%
8\times 8$ system is anomalously large compared with the infinite-size
limit. This can lead to spurious results in zero-temperature calculations.
Indeed, in Ref.\cite{Imada} it was found that the $6\times 6,$ $t^{\prime
}=\pm 0.4$ system at half-filling has long-range order at zero-temperature.
Our analytical TPSC approach suggests that long-range order is present at $%
T=0$ on both $6\times 6$ and $8\times 8$ system but not in the infinite
volume limit so that the disappearance of long-range order at zero
temperature would be apparent, in this case, only for very large systems.

\subsection{Finite-size analytical calculations}

The above considerations show that when an analytical approach\cite{Note} is
available, it can be used to understand finite-size effects by simply
comparing the infinite-size limit with calculations on systems that have the
same size as those where the simulations are done. It should also be clear
that even when an analytical approach is not available, the non-interacting
case can carry useful information on spurious finite-size effects that arise
when the thermal de Broglie wave length is larger than the system size. A
striking example of this is provided by Fig.5. At finite temperature $\left(
\beta =5\right) $, the non-interacting susceptibility has qualitatively the
same overall magnitude and wave vector dependence on a very large lattice
(Fig.5a), and on a small $8\times 8$ lattice (Fig. 5b). In the
zero-temperature limit however, Figs. 5c and 5d show that the $8\times 8$
lattice is qualitatively different from the infinite-lattice result. In
particular, on a finite lattice the maximum susceptibility is at zero wave
vector and is about five times larger than the true maximum that is
furthermore located at a totally different wave vector.

Analytical calculations on finite lattices also have difficulties of their
own in the presence of interactions. For example, the Random Phase
Approximation (RPA) can be more divergent on a finite lattice at finite
temperature than the corresponding calculation on an infinite lattice.
Indeed, consider the Lindhard function
\begin{equation}
\chi _0\left( {\bf q},iq_n=0\right) =-\frac 1N\sum_{{\bf k}}\frac{f\left(
\epsilon _{{\bf k}}\right) -f\left( \epsilon _{{\bf k+q}}\right) }{\epsilon
_{{\bf k}}-\epsilon _{{\bf k+q}}}
\end{equation}
where the sum is over the $N$ discrete lattice points of the Brillouin zone,
and $f\left( \epsilon _{{\bf k}}\right) $ is the Fermi function. Suppose
that for $m$ of the wave vectors on the Fermi surface, the condition $%
\epsilon _{{\bf k}}=\epsilon _{{\bf k+q}}$ is satisfied. Then the total
contribution of these points to the Lindhard function is given by
\begin{equation}
-\frac mN\left. \frac{\partial f\left( \epsilon _{{\bf k}}\right) }{\partial
\epsilon _{{\bf k}}}\right| _{\epsilon _{{\bf k}}=\mu }=\frac m{4TN}.
\end{equation}
On a finite lattice this contribution becomes huge at small temperatures $%
T\ll m/\left( 4N\right) $ while it is never the case in the infinite-size $%
N\rightarrow \infty $ limit. Hence, the RPA susceptibility $\chi _0/\left(
1-U\chi _0\right) $ may be divergent on a finite lattice at low temperature,
even when for an infinite system the condition $U\chi _0\ll 1$ ensuring the
validity of RPA is satisfied. Since susceptibilities are always finite on
finite systems, this shows that perturbative techniques that are appropriate
for infinite lattices may fail when applied on small systems. In the case
just discussed, the infinite-lattice result obtained from RPA may be closer
to the true finite-lattice result than the same RPA calculation done on the
finite system. Consequently, in the TPSC approach\cite{Vilk} described
below, the intermediate quantity $\chi _0\left( {\bf q},iq_n=0\right) $
appearing in the calculations sometimes has to be computed for a discrete
set of ${\bf q}$ wave vectors but with a sum over ${\bf k}$ done for the
infinite system to correctly estimate finite-size effects. Note that when
the de Broglie wave length is smaller than the system size, then the
inequality $\left( \xi _{th}=v_F/\left( \pi T\right) \right) <N^{1/2}$
ensures that the troublesome condition $T\ll m/\left( 4N\right) $ does not
occur since $T>v_F/\left( \pi N^{1/2}\right) $ generally implies that $T\gg
1/N$ is satisfied.

\section{Magnetic structure factor and Two-Particle Self-Consistent
approximation.}

The Two-Particle Self-Consistent approximation (TPSC)\cite{Vilk} agrees at
the few percent level with simulations on the nearest-neighbor model up to
intermediate coupling $(U<8)$. The agreement is overall better than FLEX or
parquet.\cite{FLEXParquet} While direct diagrammatic approaches\cite
{ChenLiBourbonnais}\cite{BulutScalapinoWhite} can work well far from
half-filling, they fail as soon as antiferromagnetic fluctuations start to
increase. Hence, we compare our Monte Carlo results only with the TPSC
approach. We first briefly discuss the approach and use it in following
subsections to understand the Monte Carlo results.

\subsection{TPSC approach}

The TPSC approach\cite{Vilk}\cite{Vilk2} can be summarized as follows. One
approximates spin and charge susceptibilities $\chi _{sp}$, $\chi _{ch}$ by
RPA-like forms but with two different effective interactions $U_{sp}$ and $%
U_{ch}$ which are then determined self-consistently. Although the
susceptibilities have an RPA functional form, the physical properties of the
theory are very different from RPA\ because of the self-consistency
conditions on $U_{sp}$ and $U_{ch}$. The necessity to have two different
effective interactions for spin and for charge is dictated by the Pauli
exclusion principle $\langle n_\sigma ^2\rangle =\langle n_\sigma \rangle $
which implies that both $\chi _{sp}$ and $\chi _{ch}$ are related to only
one local pair correlation function $\langle n_{\uparrow }n_{\downarrow
}\rangle $. Indeed, using the fluctuation-dissipation theorem in Matsubara
formalism and the Pauli principle one can write:

\begin{equation}
\frac 1{\beta N}\sum_q\chi _{ch}(q)=n+2\langle n_{\uparrow }n_{\downarrow
}\rangle -n^2=\frac 1{\beta N}\sum_q\frac{\chi _0(q)}{1+\frac 12U_{ch}\chi
_0(q)},  \label{sumCharge}
\end{equation}
\begin{equation}
\frac 1{\beta N}\sum_q\chi _{sp}(q)=n-2\langle n_{\uparrow }n_{\downarrow
}\rangle =\frac 1{\beta N}\sum_q\frac{\chi _0(q)}{1-\frac 12U_{sp}\chi _0(q)}%
,  \label{sumSpin}
\end{equation}
where $\beta \equiv 1/T$, $n=\langle n_{\uparrow }\rangle +\langle
n_{\downarrow }\rangle $, $q=({\bf q},iq_n)$ with ${\bf q}$ the wave vectors
of an $N$ site lattice, $iq_n$ Matsubara frequencies and $\chi _0(q)$ the
susceptibility for non-interacting electrons. The first equalities in each
of the above equations is an exact sum-rule, while the last equalities
define the TPSC approximation for $\chi _{ch}(q)$ and for $\chi _{sp}(q)$.
In this approach, the value of $\langle n_{\uparrow }n_{\downarrow }\rangle $
may be obtained self-consistently\cite{Vilk} by adding to the above set of
equations the relation $U_{sp}=g_{\uparrow \downarrow }(0)\,U$ with $%
g_{\uparrow \downarrow }(0)\equiv \langle n_{\uparrow }n_{\downarrow
}\rangle /\langle n_{\downarrow }\rangle \langle n_{\uparrow }\rangle .$ As
shown in Ref.\cite{Vilk}, the above procedure reproduces both
Kanamori-Brueckner screening as well as the effect of Mermin-Wagner thermal
fluctuations, giving a phase transition only at zero-temperature in two
dimensions. In general, there is however a crossover temperature $T_X$ below
which the magnetic correlation length $\xi $ grows exponentially.
Quantitative agreement with Monte Carlo simulations is obtained\cite{Vilk}
for all fillings and temperatures in the weak to intermediate coupling
regime $U<8$. The equation for charge Eq.(\ref{sumCharge}) is not necessary
to obtain the spin structure factor discussed below.

\subsection{Numerical results and interpretation of the maximum in the
uniform magnetic susceptibility}

As above, at ${\bf q}=0$ the magnetic structure factor is related by the
fluctuation-dissipation theorem to the uniform magnetic susceptibility, $%
\chi \equiv \chi _{sp}\left( q=0\right) $ which can be measured either
directly\cite{Torrance} or through NMR Knight shift. One of the puzzling
aspects of this quantity in high-temperature superconductors is the doping
dependence of its extrapolated zero-temperature value. It increases with
doping before decreasing again for larger dopings.

The uniform magnetic susceptibility $\chi $ that we obtain numerically is
plotted as a function of filling in Fig. 2 for the nearest-neighbor model
(hopping parameter $t$)\cite{MoreoSuscep}\cite{ChenSuscep} and in Fig. 3 for
the model where next-nearest neighbor hopping (hopping parameter $t^{\prime
} $) is allowed. Recalling that filling $\left\langle n\right\rangle $, that
measures the average number of electron per site, is equal to one minus
doping, the experimentally observed maximum is qualitatively described by
both models. The physical origin of both results is however very different.
In the $t^{\prime }=0$ model, Fig. 2, the maximum comes purely from the
interactions since the $U=0$ model does not have a maximum as a function of
doping. By contrast, the model $t^{\prime }\neq 0$ in Fig. 3 has a maximum
even at $U=0$ (dashed-dotted line). This maximum is simply enhanced by
interactions, as we now discuss.

In the $t^{\prime }\neq 0$ case, Fig. 3, the magnitude of finite-size
effects coming from the thermal de Broglie wavelength can be estimated from
the $U=0$ result that is plotted for both $8\times 8$ and infinite lattice.
Figs. 6a and 6b show that at all ${\bf q}$ the overall agreement is
excellent between the TPSC approach\cite{Vilk} and Monte Carlo simulations
of the structure factor. The two fillings illustrated correspond to
half-filling and to a filling close to the van Hove singularity. We have
already commented on the discrepancy for the half-filled case at ${\bf q}%
=\left( \pi ,\pi \right) $ in the section on finite-size effects. For other $%
{\bf q}$ points, the discrepancy with the TPSC is larger for the ${\bf q}=0$
results as in Fig. 3 because in TPSC it is only the integral over all ${\bf q%
}$ that is determined self-consistently. Given the difference in scale
between the results at ${\bf q}=0$ and ${\bf q}=\left( \pi ,\pi \right) $,
it is understandable that the ${\bf q}=0$ result is less precise.
Finite-size analysis of the $U=0$ case and TPSC itself (finite mesh
calculation) suggests that Monte Carlo calculations on $10\times 10$ would
suffice to completely eliminate finite-size effects. The TPSC then allows us
to conclude that the physical origin of the maximum in $\chi $ as a function
of filling for the $t^{\prime }=-0.35$ model in Fig.3 is a band-structure
effect (van Hove singularity) which is already present for the
non-interacting system $(U=0)$ and only enhanced by interactions. No
complication in the interpretation can arise from antiferromagnetism since,
as discussed in the section on finite-size effects, it does not occur in
this $t^{\prime }=-0.35$ model even at zero temperature.

By contrast, in the $t^{\prime }=0$ model of Fig. 2 the mechanism by which
interactions create the maximum is not so clear. A phenomenological
interpretation in weak coupling has been proposed.\cite{Mila} However, the
TPSC approach does not show a maximum nor does the slave-boson approach\cite
{Fresard} for this value $U=4$. This qualitative disagreement strongly
suggests that the maximum is a self-energy effect (appearance of a
pseudo-gap). This self-energy effect is important in strong coupling where
both slave-bosons\cite{Fresard} and numerical simulations\cite{MoreoSuscep}
do show a maximum. It can also be important in the presence of strong
antiferromagnetic fluctuations even at relatively weak coupling $U=4$ but it
is not included in the present version of either the TPSC approach or
slave-bosons. However, one can show that the self-energy obtained at the
next level of approximation in a manner consistent with the TPSC approach
does show a pseudogap.\cite{Vilk2} This pseudogap appears as soon as the
antiferromagnetic correlation length becomes larger than the thermal de
Broglie wavelength $\xi _{th}$. The pseudogap is more pronounced at
half-filling, leading to a relative decrease of the density of states, and
correspondingly of the magnetic susceptibility. This is what might explain
the maximum away from half-filling: As one increases the electronic density
towards half-filling, the magnetic susceptibility increases because of the
increase in density of states until the presence of antiferromagnetic
fluctuations creates a pseudogap that depletes the density of state more and
more as we move towards half-filling.

Finite-size effects on the pseudogap are negligible as long as $\xi _{th}$
is smaller than the system size, even when the antiferromagnetic correlation
length is larger than system size.\cite{Vilk2} The TPSC as well as numerical
simulations at ${\bf q=}\left( \pi ,\pi \right) $ do show that, for the
parameters of Fig. 2, we are precisely in this situation, namely the
antiferromagnetic correlation length is larger than $\xi _{th}$ but $\xi
_{th}$ is less than or of the order of the system size.\cite{ThreeBand}
Hence the apparent convergence of the Monte Carlo results as a function of
system size, despite the large antiferromagnetic correlation length, would
be consistent with the above self-energy (pseudogap) interpretation of the
maximum.

\subsection{Antiferromagnetic correlations vs zero-temperature phase diagram}

Recently, Duffy and Moreo\cite{Duffy} have found antiferromagnetism for the $%
t^{\prime }=-0.2$ model at half-filling for both $U=4$, $\beta =6$, and $U=6$%
, $\beta =4$. They also found that antiferromagnetism disappears away from
half-filling. This seems in contradiction with the zero-temperature phase
diagram\cite{TheseBenard} which predicts that antiferromagnetism would be
favored near the filling corresponding to the van Hove singularity that is
here at finite hole doping. We checked that the TPSC approach accurately
reproduces the results of Duffy and Moreo\cite{Duffy} and allows us to
understand their origin. The crossover to a regime where in two dimensions
the antiferromagnetic correlation length starts to grow exponentially\cite
{Vilk} occurs at a temperature $T_X$ that is comparable with $t^{\prime }$,
hence the details in the band structure introduced by $t^{\prime }$ are not
crucial and antiferromagnetism appears first at $n=1$, as in the $t^{\prime
}=0$ case. Note that even in this $t^{\prime }=0$ case, the TPSC predicts%
\cite{Vilk} that slightly away from half-filling the crossover to a rapidly
growing correlation length occurs at the antiferromagnetic wave vector, not
at an incommensurate wave vector, again in apparent contradiction with the
zero-temperature phase diagram. The key point in all this is that the shape
and position of the maximum in $\chi _{sp}({\bf q,}iq_n=0)$ at $T=T_X$ can
be qualitatively different from that at $T=0$. This shows that it is a
misconception to believe that the zero-temperature phase diagram determines
the high-temperature finite-size calculations, or even the real
three-dimensional finite-temperature phase transition. In fact whether the
finite-temperature phase transition is commensurate {\it or} incommensurate
may be predicted in a more realistic manner from TPSC or finite-size scaling
at finite temperature rather than from zero-temperature calculations. For
example, two-dimensional systems do not show a real phase transition until
zero temperature. Nevertheless, if finite-size calculations reveal that the
antiferromagnetic correlation length starts to grow with system size at $T_X$%
, then in real systems small three-dimensional effects would generally lead
to long-range order at the wave vector at which the two-dimensional $\chi
_{sp}$ has a maximum.\cite{DareThese}\cite{DareVilkTremblay} In the $%
t^{\prime }=0$ case slightly away from half-filling this would be at the
antiferromagnetic wave vector, not at the incommensurate one.

\section{Superconducting correlations and T-matrix effects}

The order parameter for superconductivity is defined by,
\begin{equation}
\Delta _\alpha ^{\dagger }\equiv \sum_i\Delta _\alpha ^{\dagger }\left( {\bf %
r}_i\right) \equiv \frac 1{2\sqrt{N}}\sum_{i,\nu }g^\alpha \left( \nu
\right) c_{i,\uparrow }^{\dagger }c_{i+\nu ,\downarrow }\quad ,
\end{equation}
where the sum over $\nu $ is over the nearest-neighbor sites of $i$. The
form factor determines the spatial symmetry $(s$, extended $S,p,d)$ of the
Cooper pair.\cite{ChenGap} For singlet order parameters the form factor is
even in space $g^\alpha \left( \nu \right) =g^\alpha \left( -\nu \right) $,
and the spatial symmetries usually considered include:
\begin{equation}
s\text{-wave, with }g^\alpha \left( \nu \right) =\delta _{\nu ,0}
\end{equation}
\begin{equation}
\text{extended}\,S\text{ -wave, with }g^\alpha \left( \nu \right) =\delta
_{\nu ,{\bf x}}+\delta _{\nu ,-{\bf x}}+\delta _{\nu ,{\bf y}}+\delta _{\nu
,-{\bf y}}
\end{equation}
\begin{equation}
d\text{-wave, with }g^\alpha \left( \nu \right) =\delta _{\nu ,{\bf x}%
}+\delta _{\nu ,-{\bf x}}-\delta _{\nu ,{\bf y}}-\delta _{\nu ,-{\bf y}}.
\end{equation}

Our results for the next-nearest-neighbor model $t^{\prime }=-0.35$ are
shown in Figs. 8 and 9. Size effects are still quite large. At this
temperature, we cannot really expect results which would converge to the
infinite size limit before $10\times 10$ lattices. To account for this, the
calculations for the $T$-matrix are done for systems of the same size as the
Monte-Carlo simulations, namely $8\times 8$ or $6\times 6$.

For the nearest-neighbor model $(t^{\prime }=0)$, we have shown in an
earlier paper\cite{Dare} that most of the correlations at low density can be
accounted for by two-body particle-particle scattering, namely $T-$matrix
effects. Near half-filling, one observes deviations from $T-$matrix
predictions, especially for the $d$-wave correlations. Fig. 7, reproduced
from Ref.\cite{Dare}, shows that the $T$-matrix underestimates
superconducting fluctuations at half-filling. Since for $d$-wave, the $T$%
-matrix gives in fact the same result as the non-interacting case\cite{Dare}%
\cite{DareThese} this means that the correlations measured by Monte Carlo
are enhanced compared with the non-interacting case. This could indicate
that either the van Hove singularity or the nesting-induced
antiferromagnetism increase $d$-wave correlations.

The $t^{\prime }=-0.35$ Monte Carlo results in Figs. 8 and 9 can help
disentangle the influence of van Hove singularity and nesting. For these
figures, we learn that: a) As expected in general and as found in particular
in the $t^{\prime }=0$ case, the $T-$matrix approximation works best at low
electron or hole filling, {\em i.e.} near $\left\langle n\right\rangle =0$
and $\left\langle n\right\rangle =2.$ b) The disagreement between
simulations and $T-$matrix is largest on the $\left\langle n\right\rangle <1$
side in the vicinity of the van Hove singularity. c) The $T-$matrix
everywhere overestimates the superconducting correlations. Even though this
should in no way be considered a proof, this last point suggests that the
finite $t^{\prime }$ model is even less favorable to superconductivity than
the nearest-neighbor model. Earlier studies,\cite{Moreo}\cite{Lin}\cite
{DosSantos}\cite{Husslein} are equally divided on this point. Here, no sign
of superconductivity is found in the size dependence of correlations
functions. A plot of the spatial dependence\cite{Moreo}\cite{Husslein} would
however be a more sensitive test.

As mentioned in the context of the magnetic structure factor, the parameters
studied here do not correspond to a regime where antiferromagnetism has
started to set in at half-filling, contrary to the studies of the $t^{\prime
}=0$ model. For the $t^{\prime }=0$ model the only case where the $T-$matrix
clearly underestimates the superconducting correlations is for $d$-wave
pairing near half-filling\cite{Dare}\cite{DareThese} at a temperature where
nesting and antiferromagnetic correlations are strong. Since the van Hove
singularity located at the same filling is expected to decrease pairing
correlations, as in remark b) above, this seems a clear indication that
antiferromagnetism increases superconducting fluctuations. We have not found
however yet that the increase is sufficient to lead to a superconducting
phase transition. To confirm the positive influence of antiferromagnetism on
superconductivity, it would be interesting in the $t^{\prime }\neq 0$ case
to either increase $U$ or decrease temperature to reach a regime where
antiferromagnetism starts to show up in the model. One should beware however
that antiferromagnetic correlations can have a positive influence on
superconducting fluctuations due simply to a short-range effect, as argued
in Ref.\cite{Scalapino}.

\section{Conclusion}

We have studied the next-nearest-neighbor Hubbard model for $t^{\prime
}=-0.35$ and $t^{\prime }=0$ using both Monte Carlo simulations and
analytical approaches. We have also discussed at length finite-size effects.
The TPSC approach\cite{Vilk} allowed us to understand the magnetic
fluctuation properties, including the occurence of antiferromagnetism
observed in previous calculations at half-filling.\cite{Duffy}

The uniform magnetic susceptibility obtained with the $t^{\prime }=-0.35$
model has the same qualitative doping dependence as the experimental results
on high-temperature superconductors. Since we have found that the TPSC
approach gives a satisfactory explanation of the spin-spin correlation
functions computed by Monte Carlo simulations, we were able to identify the
maximum in the magnetic susceptibility as mostly a band-structure effect
enhanced by interactions. We contrasted this with the $t^{\prime }=0$ model
where a maximum also shows up but for apparently very different reasons
related probably to self-energy (pseudogap) effects which can occur either
at strong coupling or in the presence of large antiferromagnetic correlation
lengths.

For the interaction and temperature ranges studied here, superconductivity
is less favored for $t^{\prime }=-0.35$ than for the $t^{\prime }=0$ model
since the Monte Carlo results are consistently smaller than trivial $T$%
-matrix effects. However, antiferromagnetic correlations do not occur in
this parameter range. It is interesting to note that despite the fact the $T$%
-matrix {\em overestimates} superconducting correlations in the presence of
a van Hove singularity, in the $t^{\prime }=0$ case the presence of nesting
and concomitant large antiferromagnetic correlations at half-filling suffice
to overcome the detrimental effect of the van Hove singularity to increase
the superconducting $d$-wave fluctuations {\em above} the $T$-matrix
prediction. This clearly suggests that antiferromagnetic correlations can
have a positive influence on $d$-wave pairing. However, the effect observed
to date is clearly a short-range one so that the issue of long-range
superconducting order caused by antiferromagnetic correlations is not
settled yet.

We acknowledge the support of the Natural Sciences and Engineering Research
Council of Canada (NSERC), the Fonds pour la formation de chercheurs et
l'aide \`{a} la recherche from the Government of Qu\'{e}bec (FCAR) and
(A.-M.S.T.) the Canadian Institute of Advanced Research (CIAR) and the
Killam foundation. A. Veilleux would like to thank the Fujitsu corporation
for a 1993 HPC/Fujitsu scholarship.

\figure Figure 1: Average sign of the determinant as a function of filling
for the next-nearest-neighbor hopping model.

\figure Figure 2: Maximum in the uniform magnetic susceptibility for the
nearest-neighbor model for various system sizes and for the TPSC approach.
In the latter case, the calculation is for an infinite system. The curve for
$4\times 4$ system is from Ref.\cite{ChenSuscep}.

\figure Figure 3: Maximum in the uniform magnetic susceptibility for the
next-nearest-neighbor model for $8\times 8$ system and for the TPSC approach
on an infinite lattice. $U=0$ results are also presented for $8\times 8$ and
for infinite size to estimate finite-size effects.

\figure Figure 4: Magnetic structure factor at the antiferromagnetic
wavevector, compared with the TPSC approach in the infinite-size limit. The
antiferromagnetic correlation length for the next-nearest-neighbor model is
largest at half-filling but remains finite for the parameters considered
here.

\figure Figure 5: Finite-size effects in the non-interacting uniform
susceptibility when $t^{\prime }=-0.35$. a) and b) calculations are at $%
\beta =5$ for, respectively, a converged lattice and an $8\times 8$ lattice.
c) and d) are for the low temperature limit $\beta =100$.

\figure Figure 6: Magnetic structure factor measured along the sides of the
irreducible Brillouin zone illustrated in the insets. The TPSC calculation
is for the infinite lattice. a) Half-filled case b) Filling near the van
Hove singularity.

\figure Figure 7: Reproduced from Fig. 6 of Ref.\cite{Dare}. Singlet $d-$%
wave corrrelation function plotted as a function of filling in the
nearest-neighbor $\left( t^{\prime }=0\right) $ model for $U=4$, $\beta =6$.
The $T-$matrix result is the same as the $U=0$ result and is calculated on
the same lattice sizes at the same temperature $\beta =6$. a) $4\times 4$
lattice, b) $8\times 8$ lattice.

\figure Figure 8: Order parameter correlations for extended $S$-wave
superconductivity. a) $8\times 8$ system for $\left\langle n\right\rangle
\leq 1$, b) $6\times 6$ system for the whole range of fillings.

\figure Figure 9: Order parameter correlations for $d$-wave
superconductivity. a) $8\times 8$ system for $\left\langle n\right\rangle
\leq 1$, b) $6\times 6$ system for the whole range of fillings.


\begin{references}
\bibitem{Note}  ''Analytical'' approaches such as perturbation theory, $T-$%
matrix or TPSC approach, in general need the help of numerical integrations.
Nevertheless, we call these approaches ''analytical'' in the sense that
their approaximations and physical interpretation follow from analytical
arguments, even though their quantitative numerical predictions can
generally be found only with the help of a computer.
\end{references}
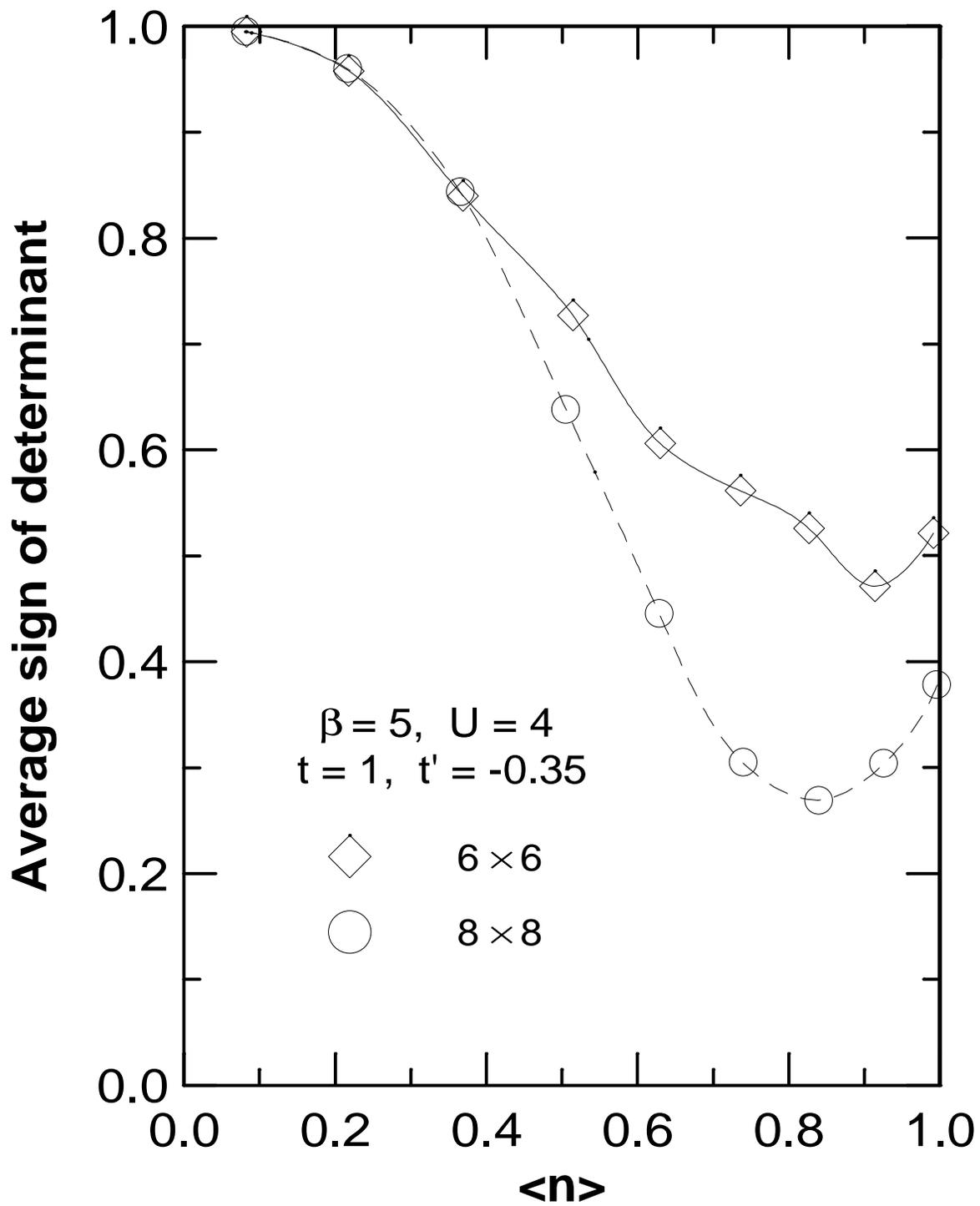
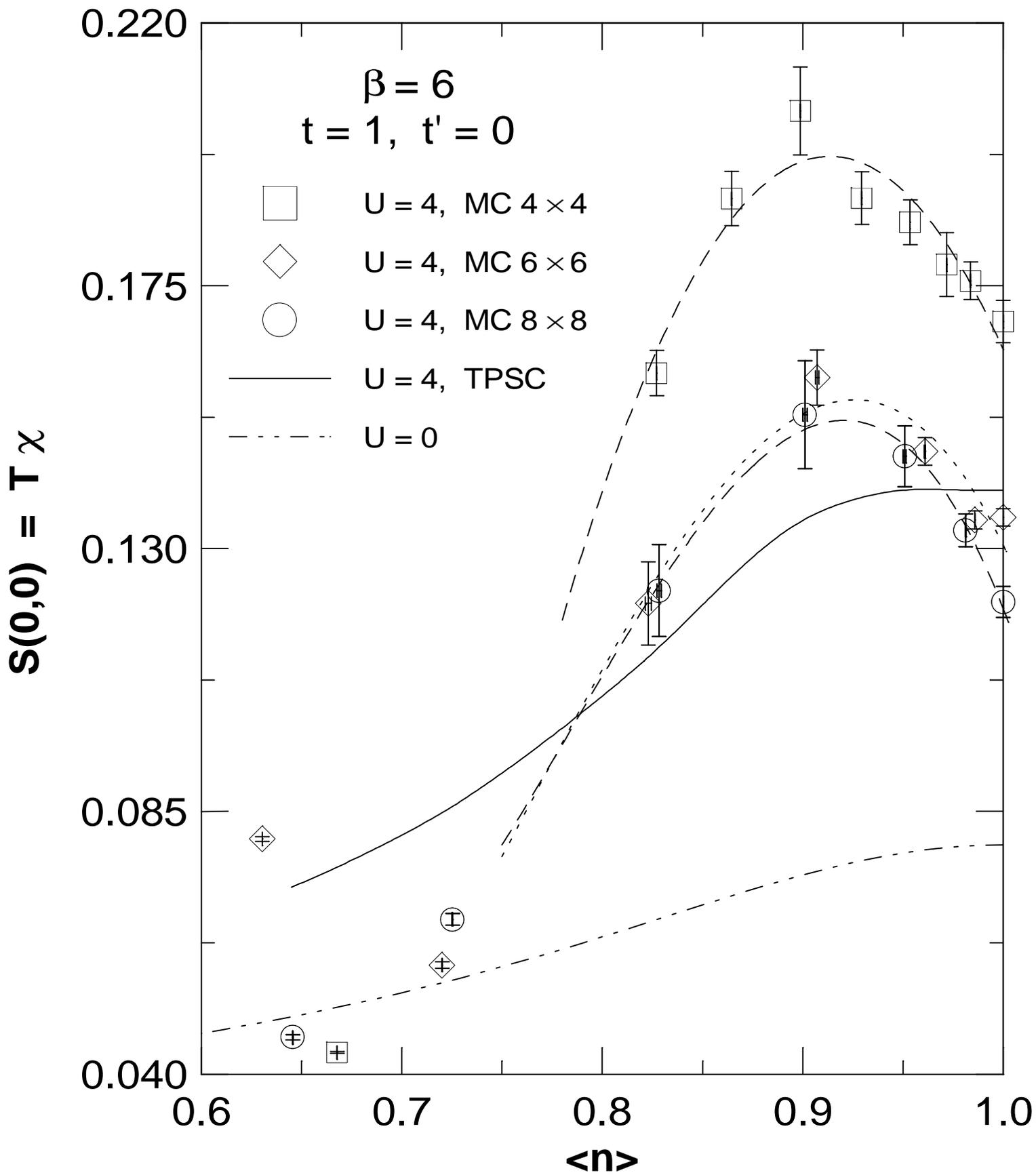
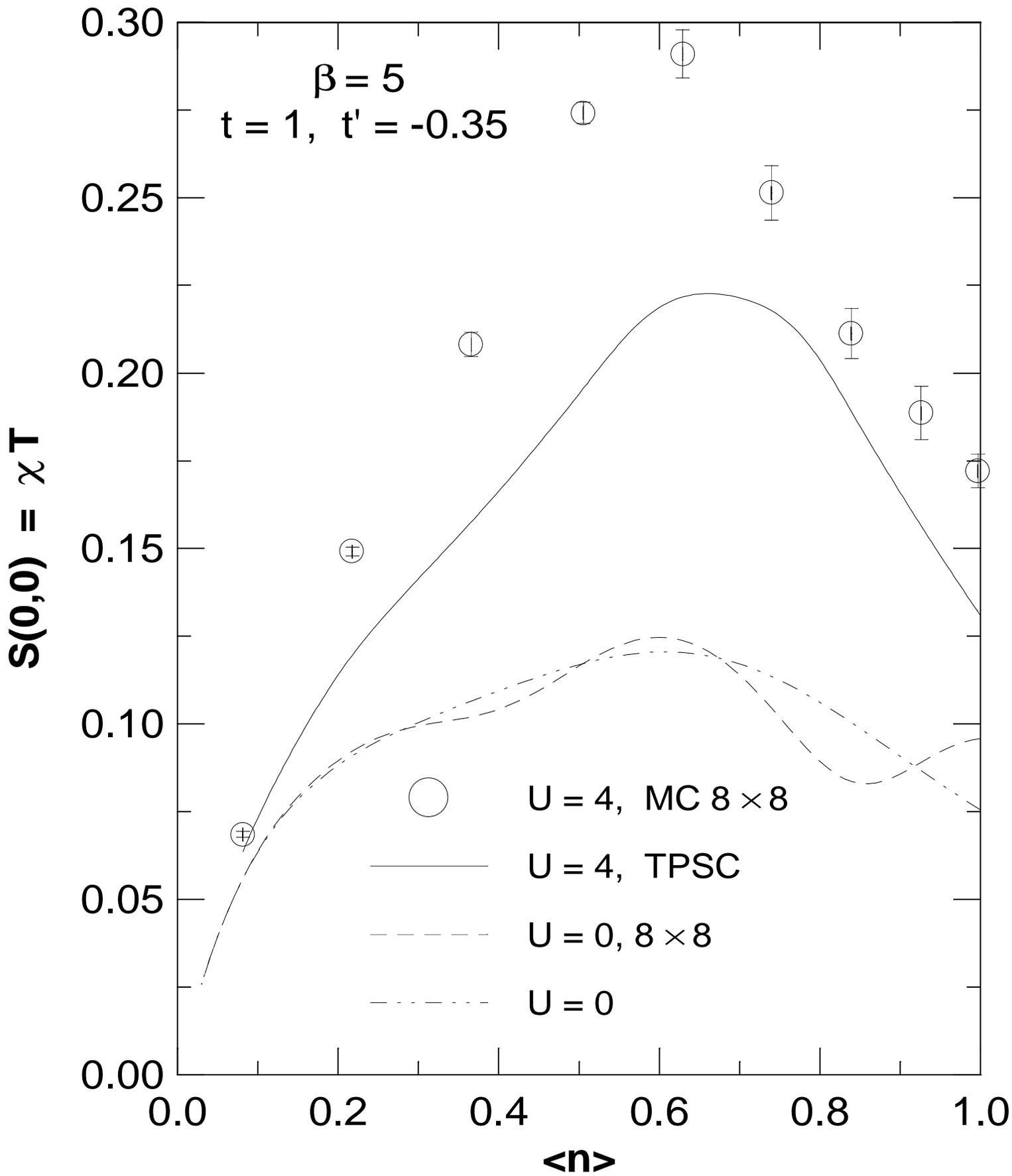
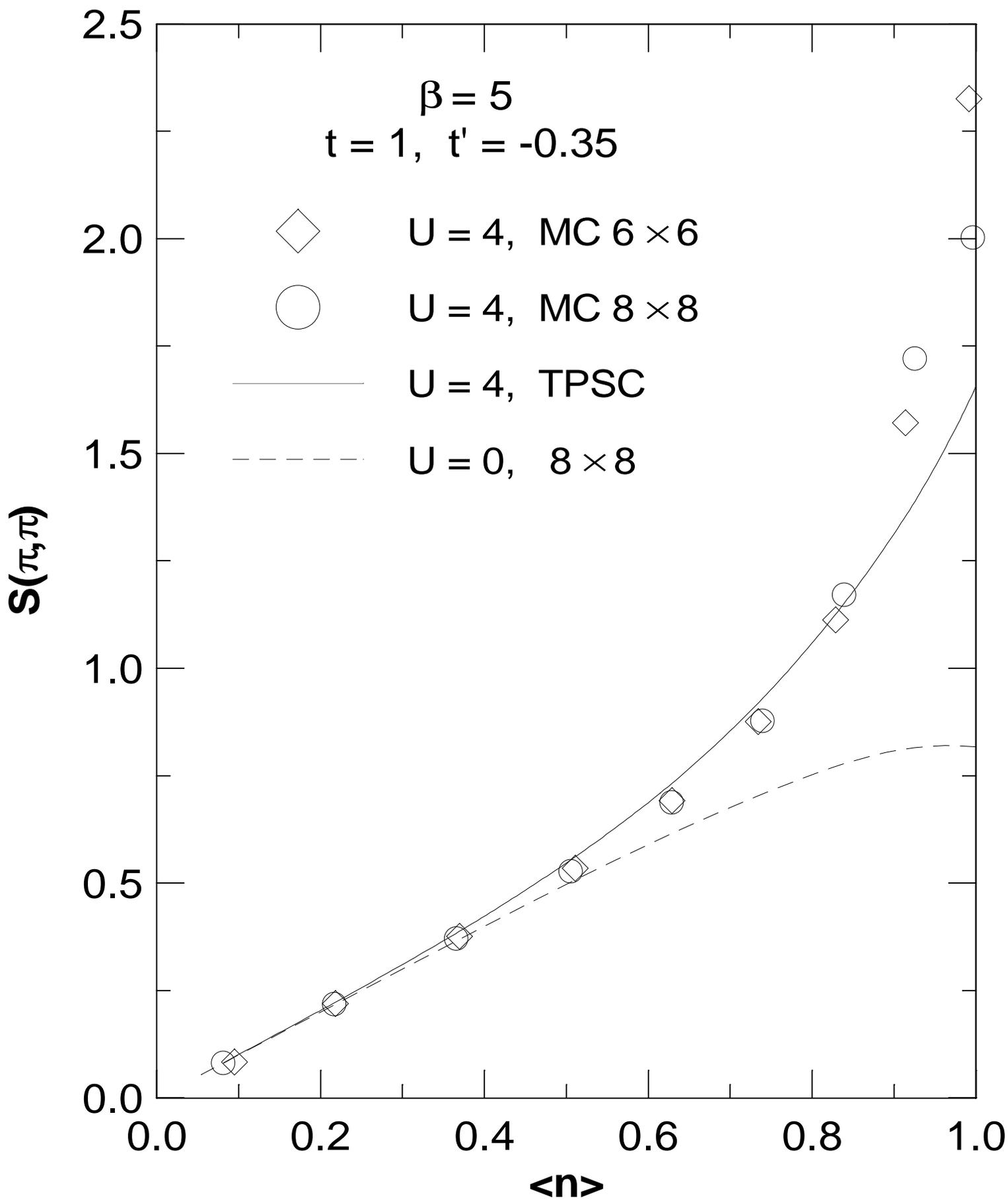
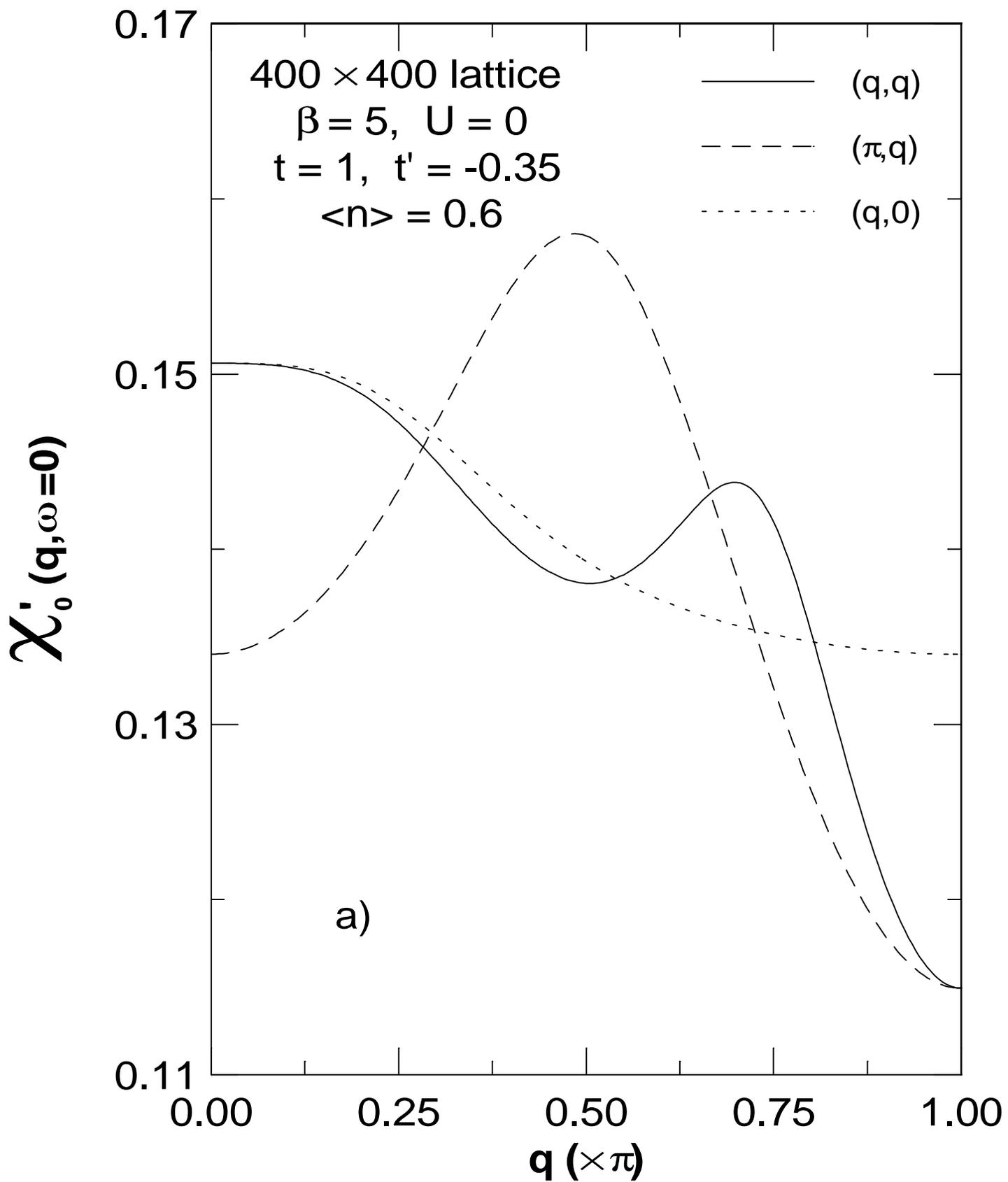
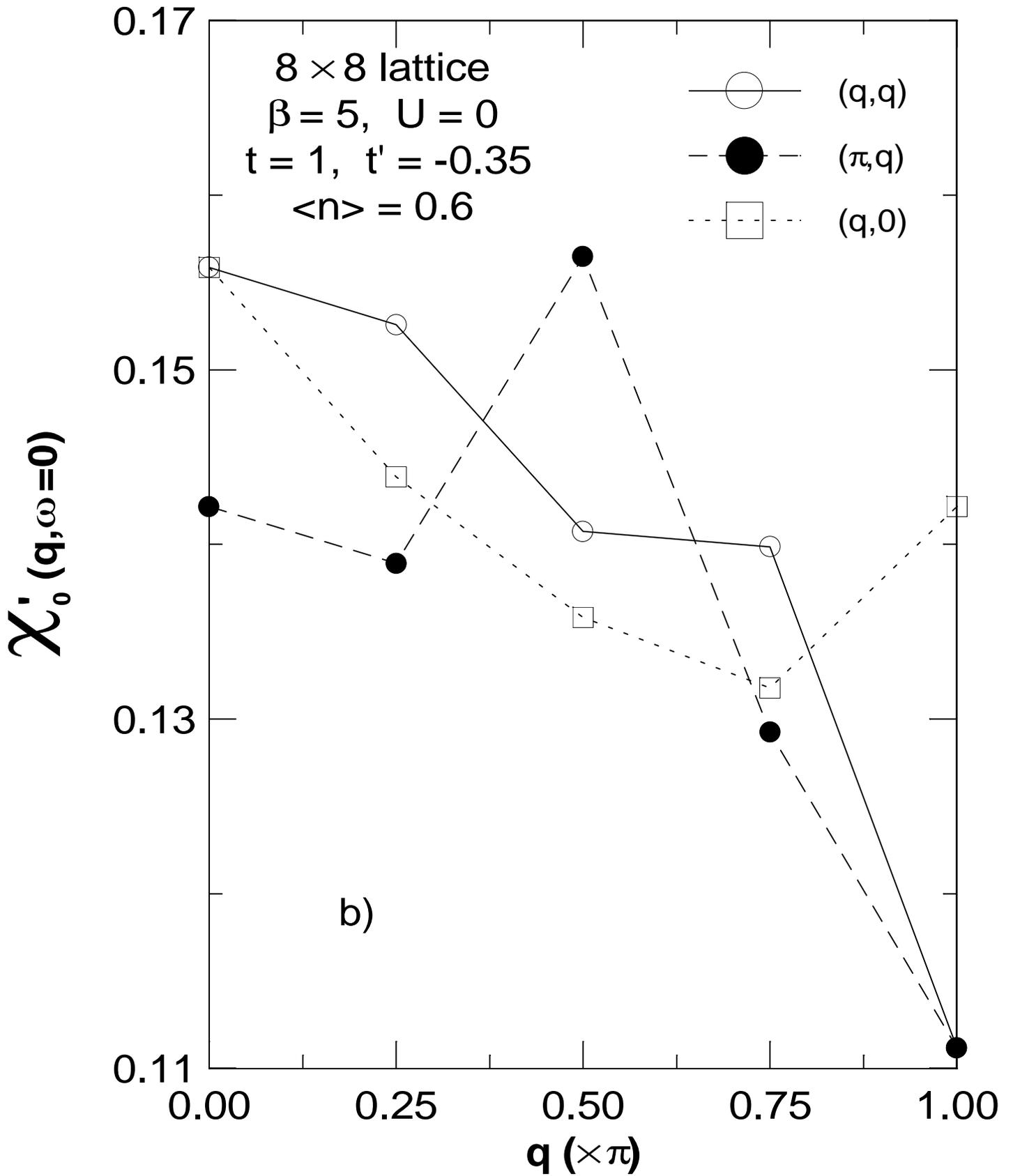
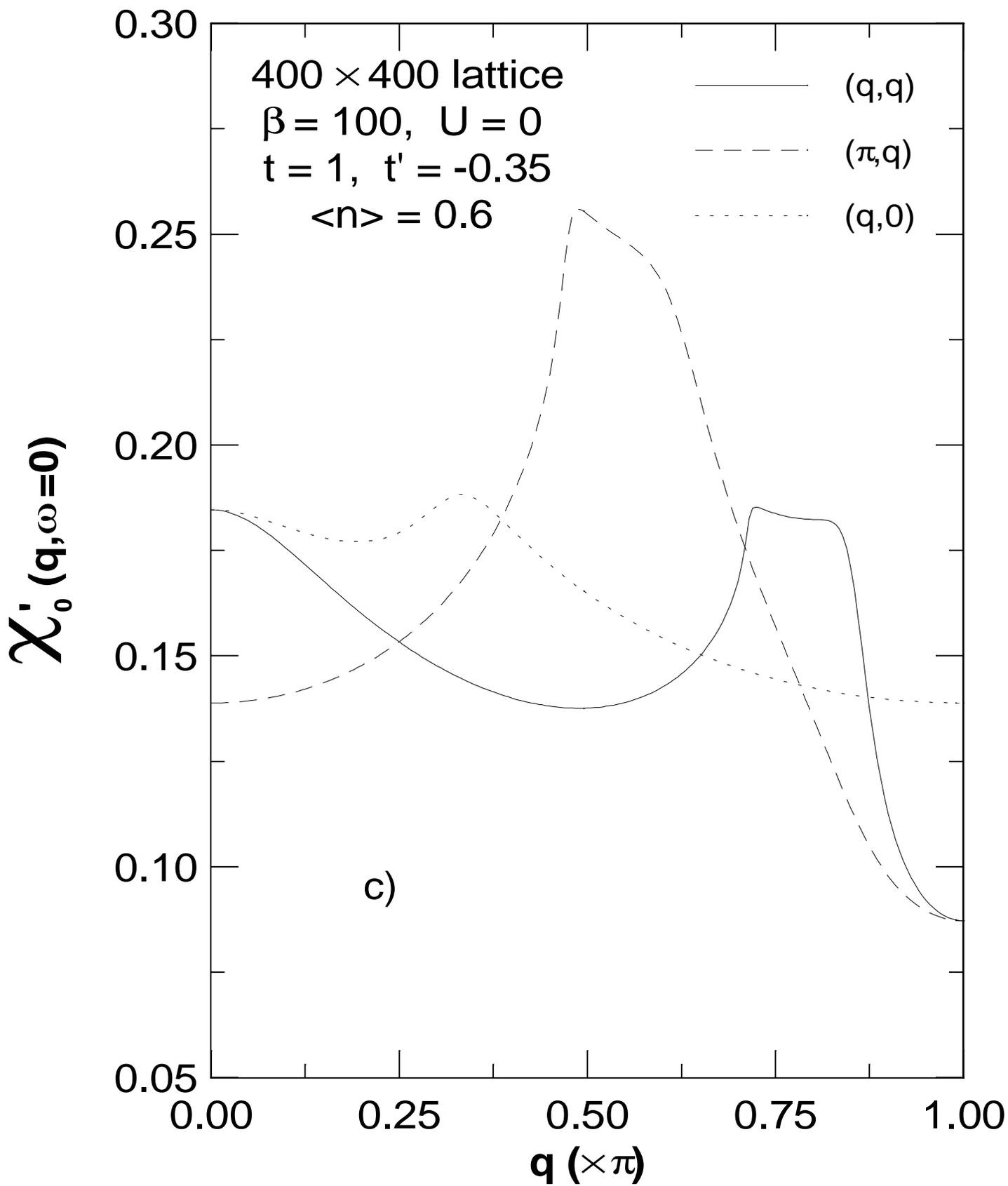
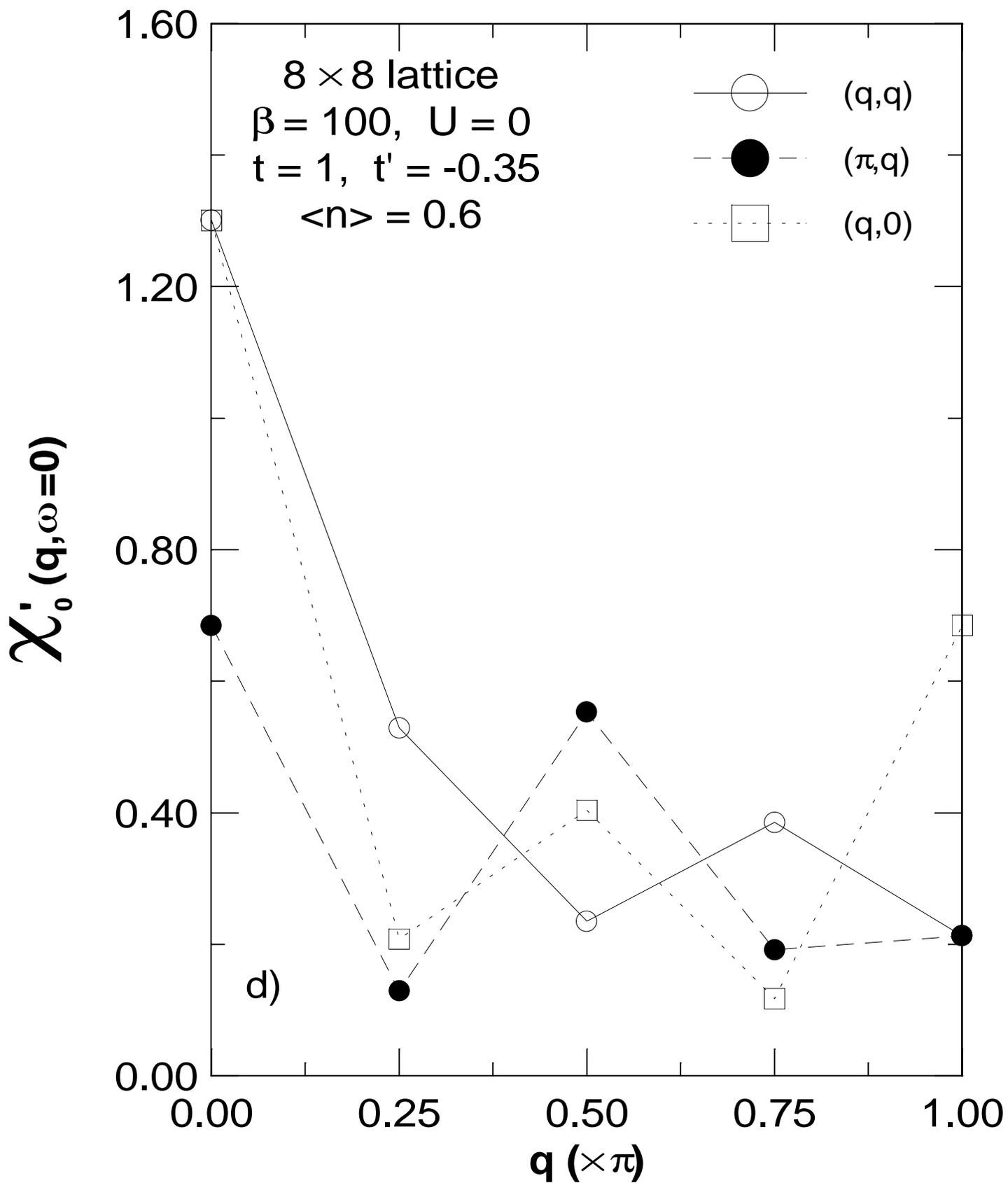
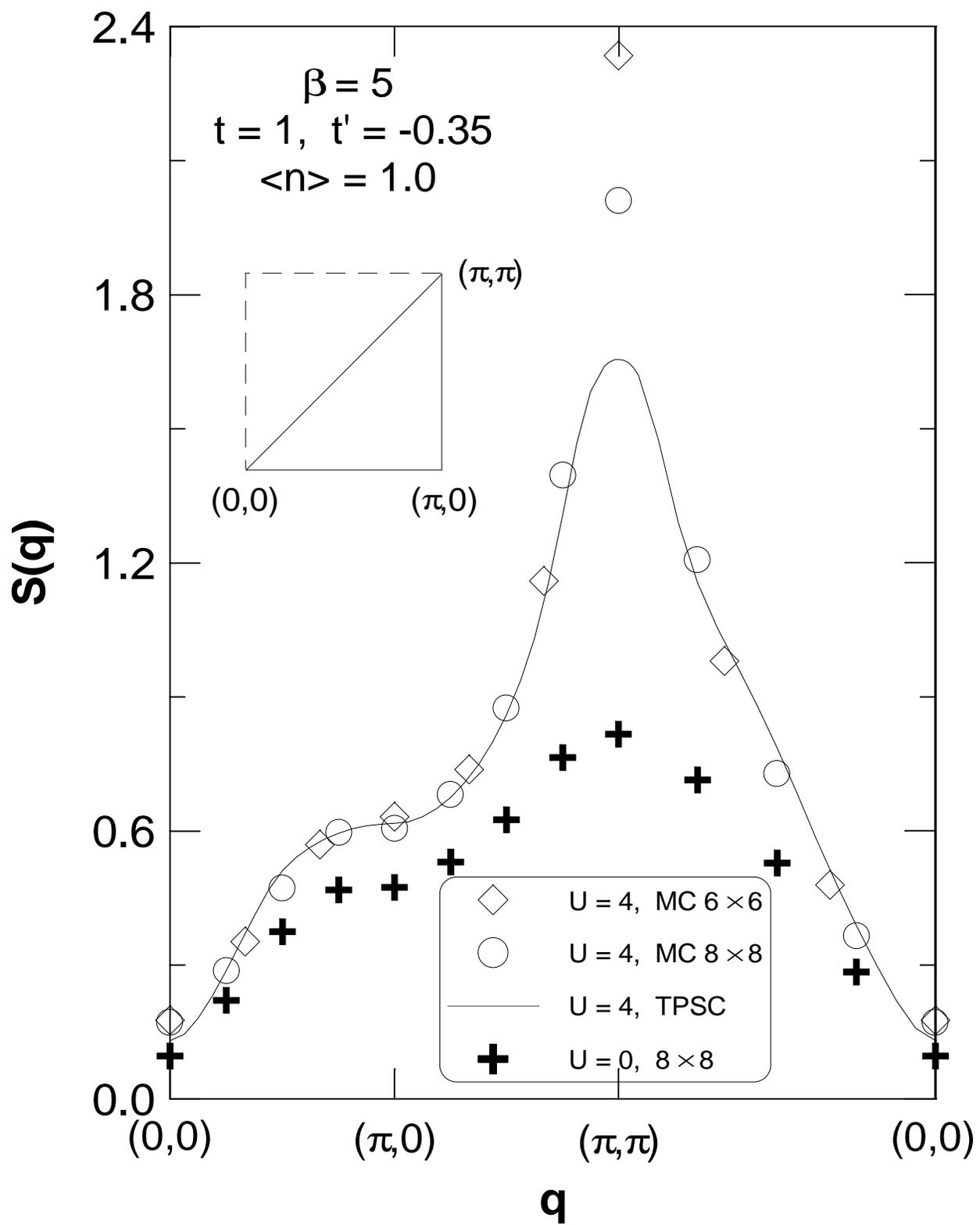
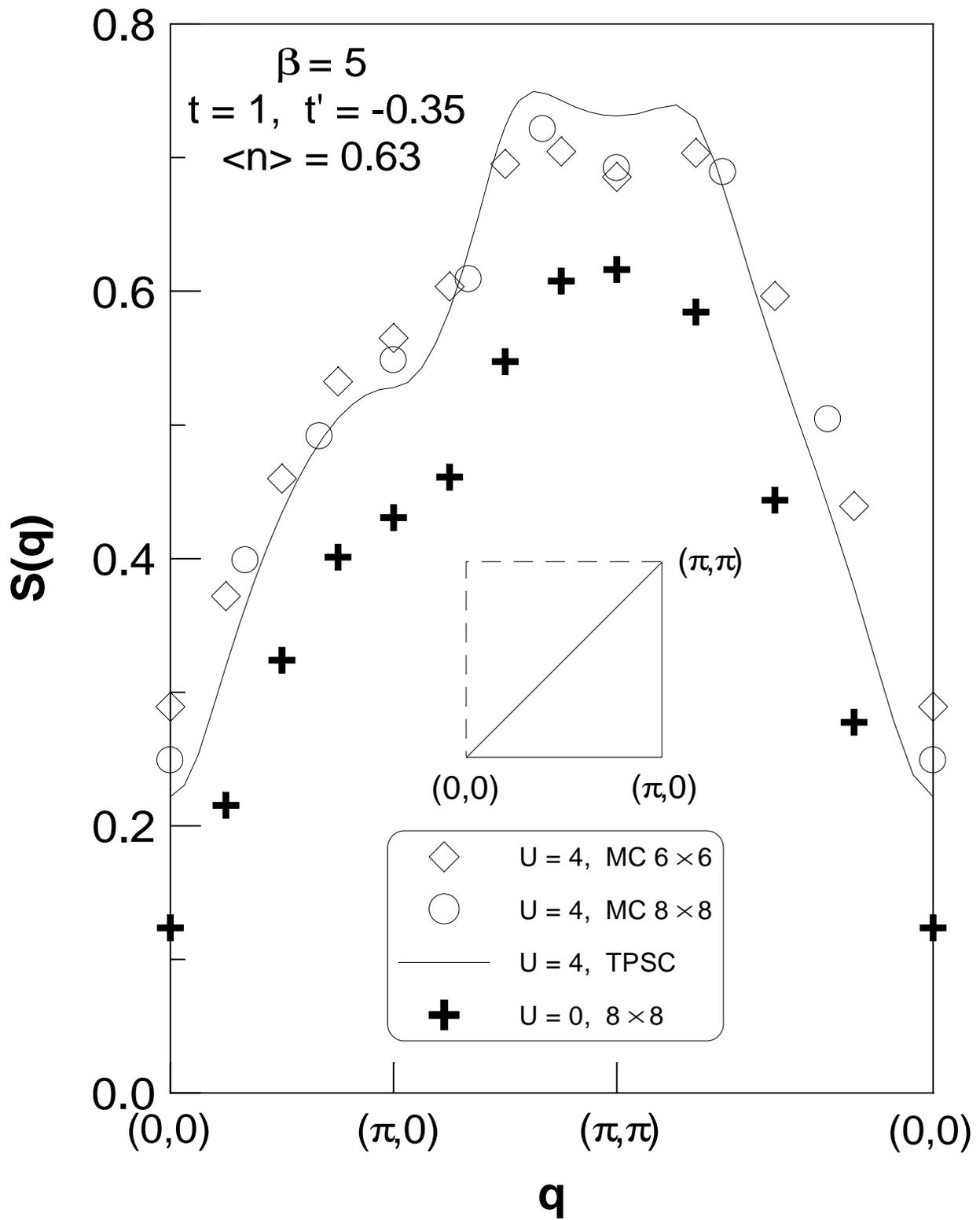
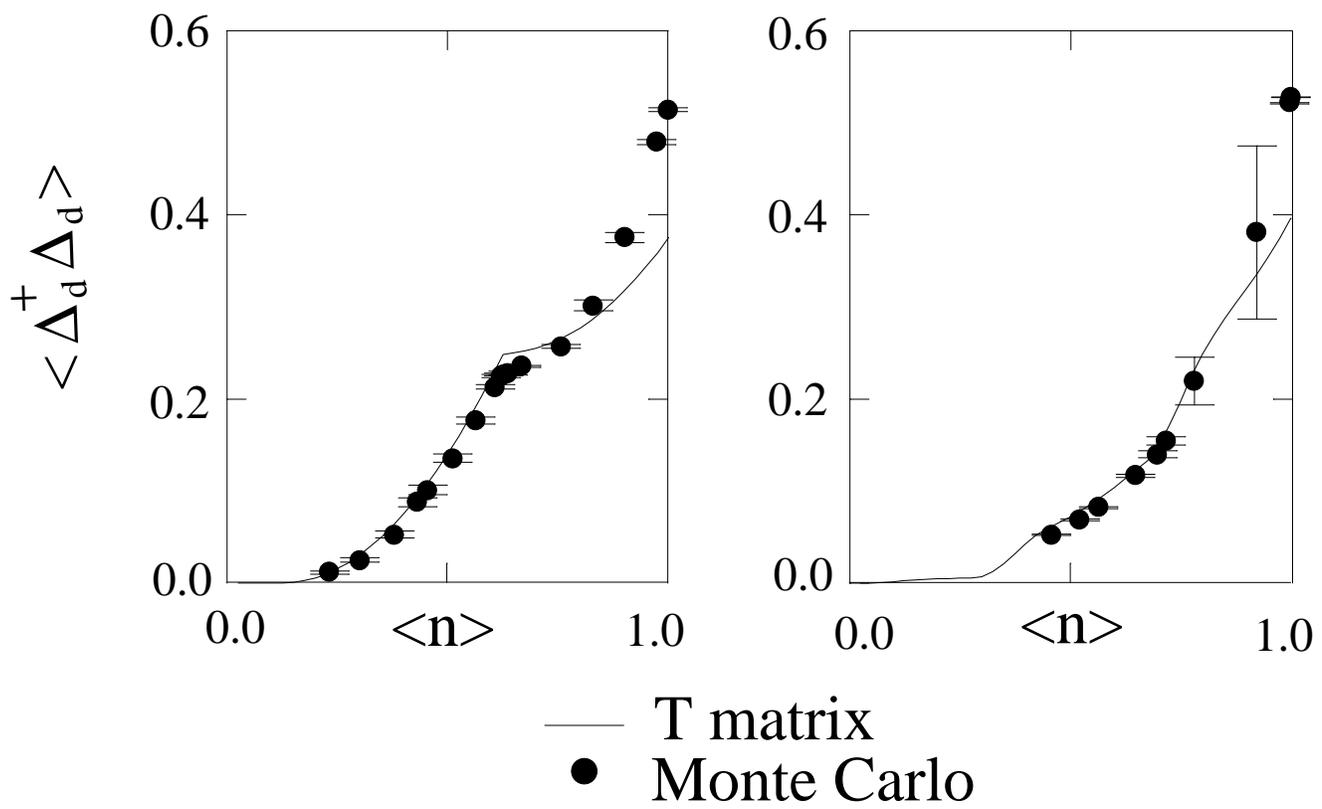
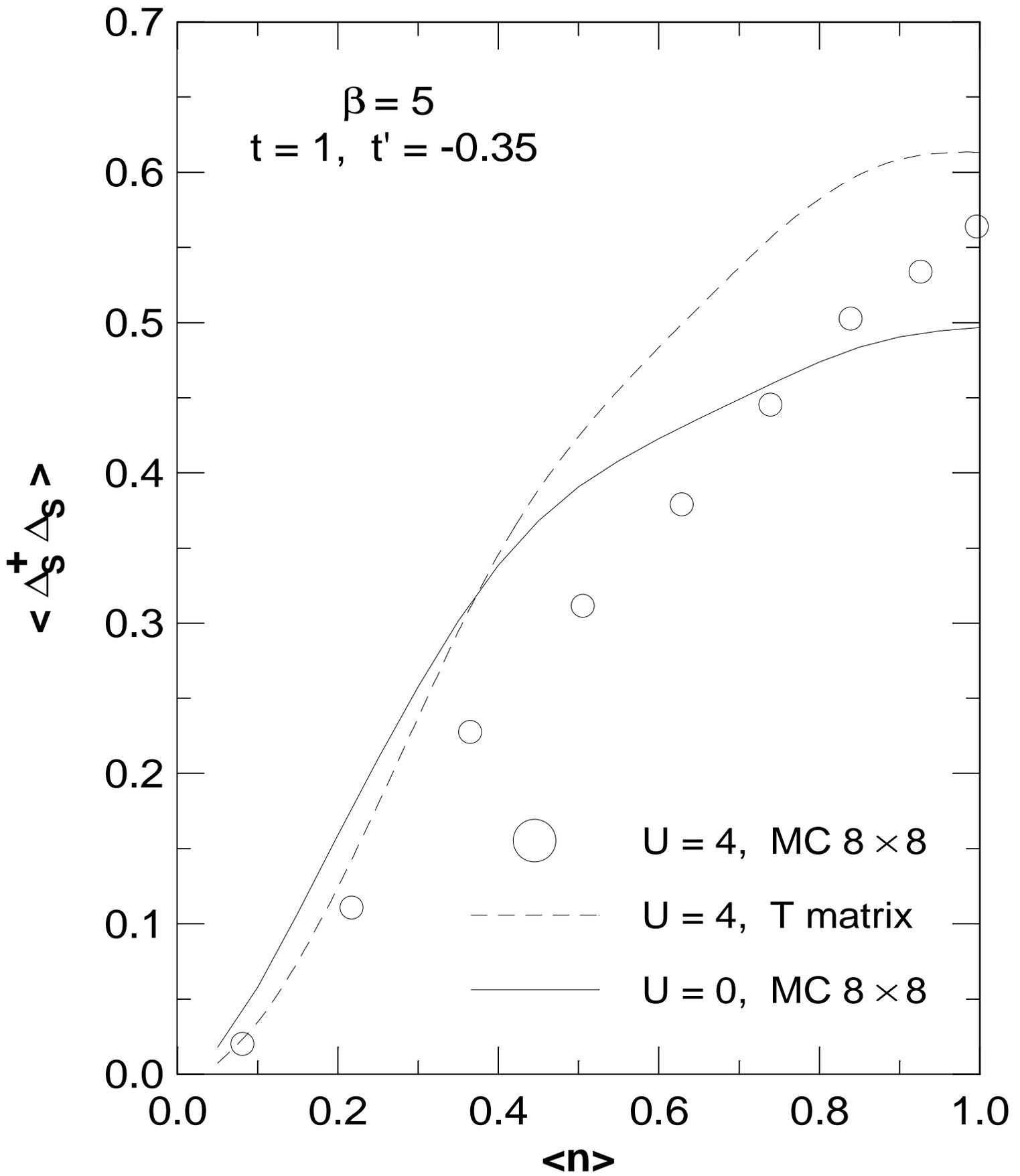
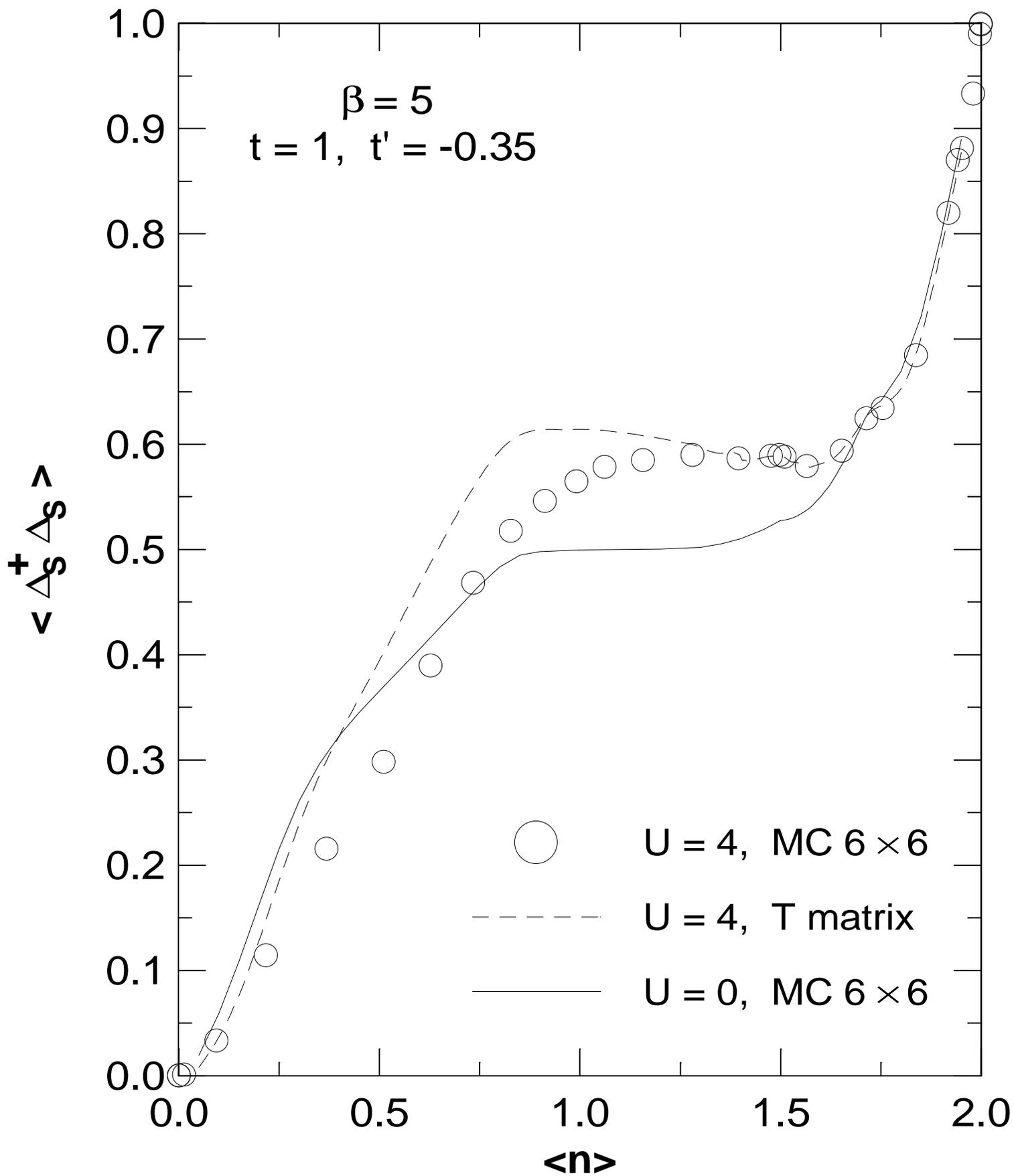
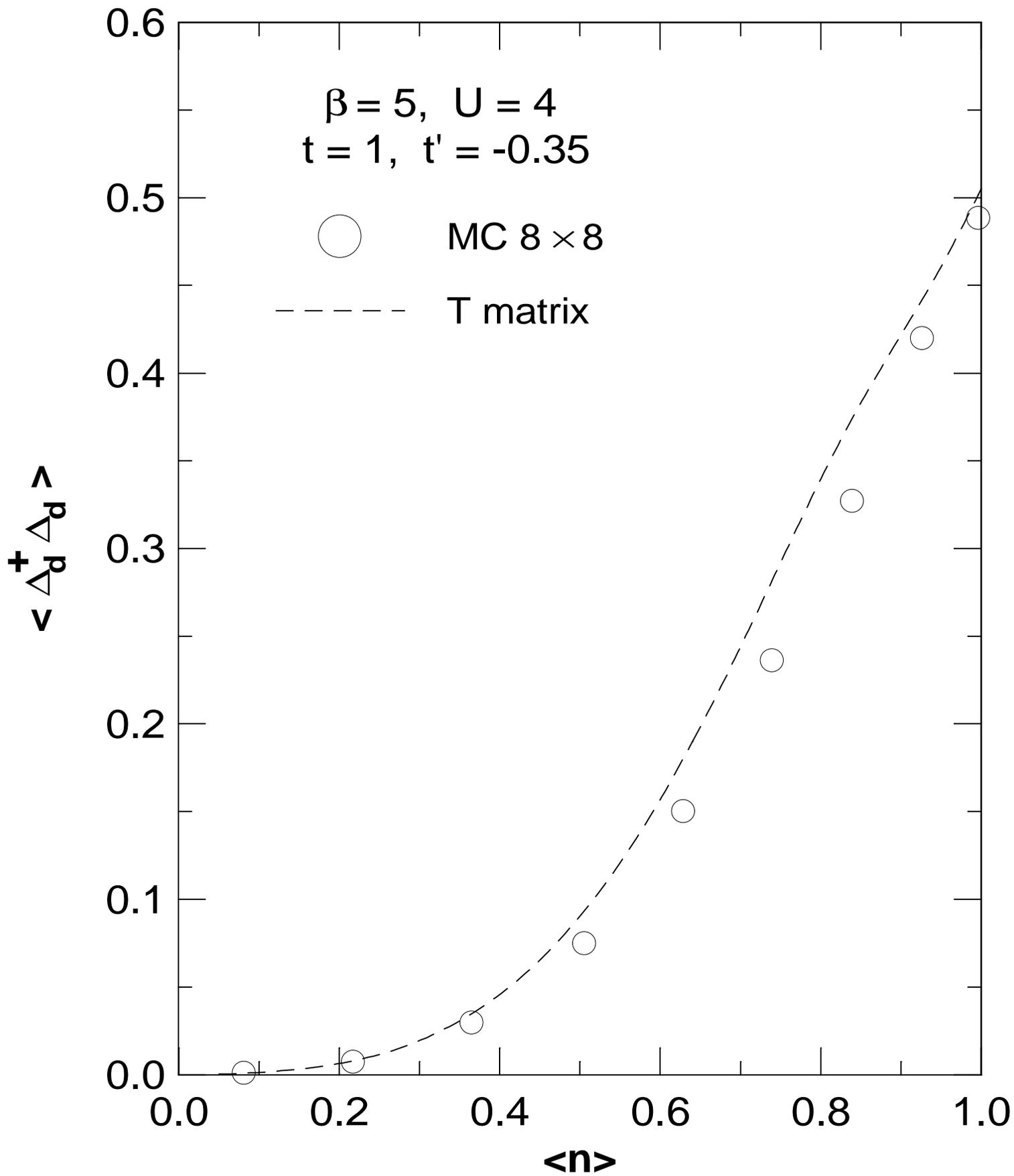
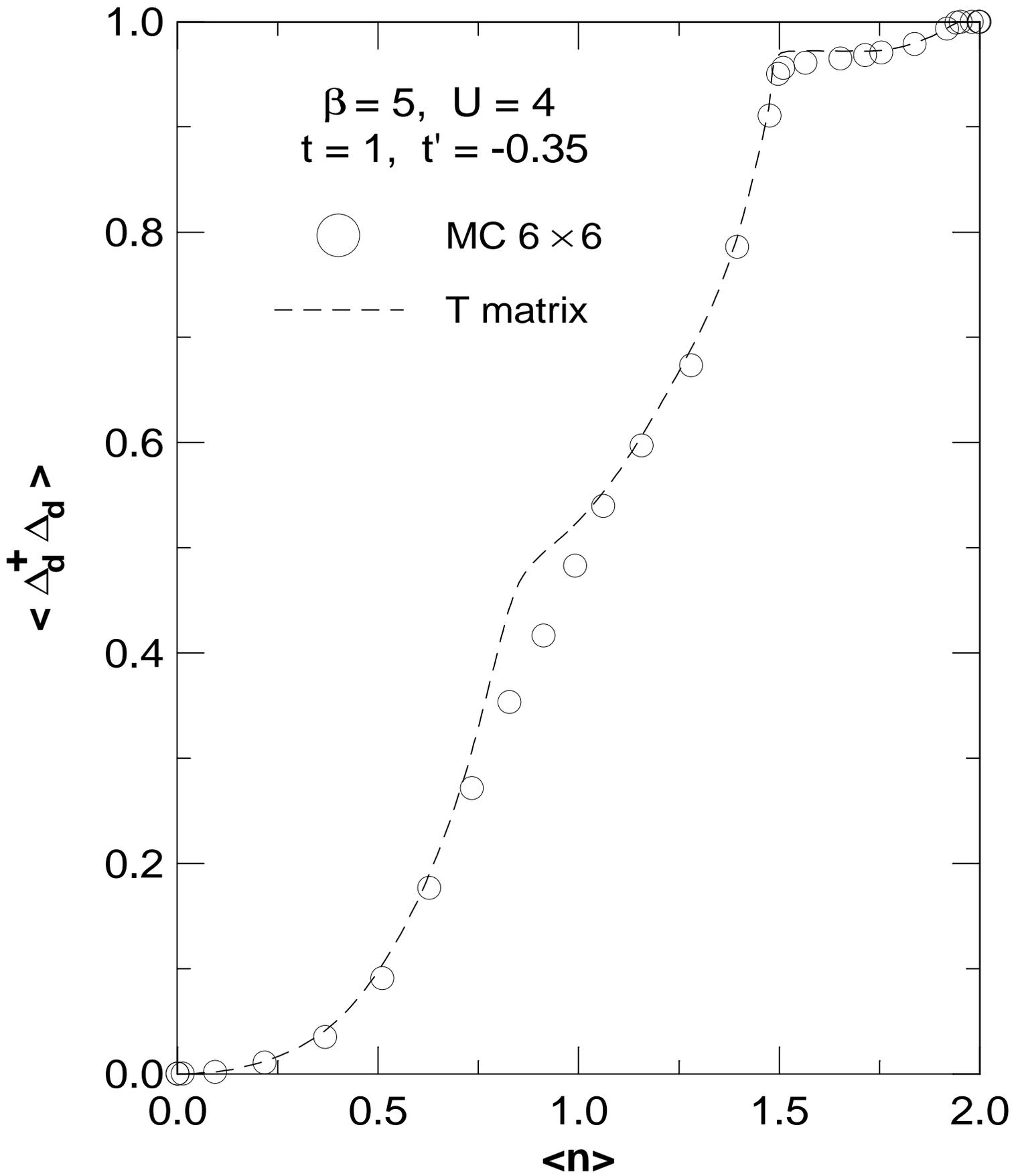
\end{document}